\documentclass[MSNbibl,nameyear,dvips]{arxstspdf}
\usepackage{flushend}
\usepackage{stfloats}

\volume{29}
\issue{3}
\pubyear{2014}
\firstpage{375}
\lastpage{379}
\doi{10.1214/14-STS496} 
\referstodoi{10.1214/14-STS480}
\docsubty{FLA}



\begin{document}
\begin{frontmatter}

\vspace*{12pt}\title{Rejoinder}
\runtitle{Rejoinder}

\begin{aug}
\author[a]{\fnms{Guido}~\snm{Imbens}\ead[label=e1]{imbens@stanford.edu}\ead[label=u1,url]{http://www.gsb.stanford.edu/users/imbens}}
\runauthor{G. Imbens}

\affiliation{Stanford University}

\address[a]{Guido W. Imbens is the Applied Econometrics Professor and Professor of Economics, Graduate School of Business,
Stanford University, Stanford, California 94305, USA and NBER \printead{e1,u1}}
\end{aug}


\end{frontmatter}

I am very grateful for the comments on the paper and the careful
reading that went into them. Since instrumental variables concepts and
methods have become popular in a range of substantive areas beyond
economics, there have been a number of significant contributions from
other areas, and it is useful to have the different perspectives on
these methods that these comments reflect.
I will attempt to address some of the issues raised in the comments,
but many of these comments will undoubtedly stimulate new studies, as
the general area of research on causal inference in observational
studies continues to flourish.

\section*{Kitagawa: ``Instrumental Variables Before
and Later''}

I am grateful for the kind words by Kitagawa. He has been doing very
interesting work on
testing for validity of instrumental variables in recent years (e.g., Kitagawa, \citeyear{Kit10}, \citeyear{Kit13}) that will undoubtedly be influential in
the literature.
I am also glad that Kitagawa likes my summary of the differences
between econometric and statistical approaches to causality as ``choice
versus chance.''

Kitagawa's comments on the impact of the local average treatment effect
literature on economic practice agree with my views. As emphasized in
the paper, the LATE concept was never intended to change the question
of interest, but to clarify what we could learn from the data.
Nevertheless, in some cases the LATE may well be representative of a
subpopulation that is of substantial interest on its own.
Consider the draft lottery example (Angrist, 1990; Hearst, Newman and
Hully, 1986) where the compliers are the men who served, or would have
served, in the military, because of their draft lottery number.
Arguably, this is the group on the margin for whom the effect of
military service is most interesting.
Similarly, in the Angrist and Krueger (1991) study of the returns to
education using compulsory schooling laws as an instrument, the
compliers are the individuals for whom schooling decisions are affected
by compulsory schooling laws, again arguably an interesting
subpopulation for educational policies that are often targeted at those
receiving lower levels of education. Nonetheless, in general the
subpopulation of compliers is not chosen for its interest, but because
we can hope to learn something about them. It is about the primacy of
internal validity over external validity (Shadish, Cook and Campbell, \citeyear{ShaCamCoo02}).

Kitagawa discusses instrumental variables in the context of another
example that, like the supply-and-demand example I discuss in the
paper, is a classic one, that of the estimation of returns on inputs in
a production function. Specifically, he focuses on the causal effect of
labor inputs on output. The starting point for an economist is exactly
as Kitagawa describes: firms do not choose input levels randomly, but
choose them optimally, for example,  to maximize profits.
This leads quickly to settings where we cannot simply regress output on
inputs if we are interested in the causal effect of input on output.
Moreover, the context in combination with economic theory on firm
behavior suggests where a researcher might look for instruments that
satisfy the exclusion restriction, namely cost variables that affect
the choice of input levels but that affect output only through their
effect on input levels.

In his comments, Kitagawa also distinguishes between various objectives
for the researcher. If the goal of the researcher is what he calls
``scientific reporting,'' Kitagawa agrees with my recommendation to
report both estimates of the local average treatment effect and bounds
on the overall average treatment effect. If, on the other hand, the
goal is directly to make a decision, say, on whether to extend the
treatment to the entire population or not, he advocates a decision
theoretic approach, either Bayesian along the lines of Chamberlain
(\citeyear{C11}), or the type of Manski ``data-alone'' frequentist approach. I
agree with that, and I think the distinction between scientific
reporting and decision making is a useful one to bear in mind.

\section*{Richardson and Robins:  ``ACE Bounds; SEMs with
Equilibrium Conditions''}

Richardson and Robins make two sets of comments, one about bounds on
the average causal effect (ACE), and one about simultaneous equations
models (SEMs).

In the discussion on bounds, they formulate four sets of assumptions,
captured by different graphical models that allow for construction of
the same set of bounds. They relate these assumptions to their novel
Single World Intervention Graphs (SWIGs).
I find the SWIGs an intriguing approach, and one that might help make
the graphical approach more relevant for researchers interested in
causal effects.
One concern I have with the discussion of the four sets of assumptions
is that it is not clear when there is a substantively important
difference between the assumptions. For example, I find it difficult to
think of substantive applications where the independencies hold one
pair at a time [Assumption (iii)], but not joint independence
[Assumption (i)].

The discussion on market equilibrium and bicausal models is very
interesting and stimulating.
I am happy to see Richardson and Robins endorse my interpretation of
structural equations in terms of potential outcomes. Although, as the
authors point out, this interpretation of structural equations is not
universal, in my view, partly based on conversations with other
economists, it is the leading one in economics. The discussions of
normalization issues that the authors refer to generally arise in the
context of estimation in settings where there are multiple instruments.
In that case, the difference between estimation methods such as limited
information maximum likelihood (LIML, going back to Anderson and Rubin,
1948), and two stage least squares (TSLS) matter. In the recent
literature on weak instruments, these differences have been shown to
potentially matter a great deal. Staiger and Stock (1997) is a key
paper, and Stock and Andrews (2005) provide an overview.

Although economists routinely use the supply-and-demand example in
textbooks and teaching, most discussions no longer explicitly discuss
where the equilibrium that is assumed arises from, making the work in
this area more difficult to access for researchers from other areas
than it need to be. The model used by Richardson (\citeyear{R96}) where the data
come from a discrete approximation to a finer recursive model appears
to capture well the mechanisms researchers implicitly have in mind. See
\citet{Ber66} for a related discussion in the older economics
literature discussing the relationship between nonrecursive (bicausal)
models in discrete time and recursive continuous time models.

\section*{Shpitser:
``Causal Graphs: Addressing the
Confounding Problem Without
Instruments or Ignorability''}

Shpitser is concerned that I did not discuss the growing literature on
causal graphical models.
This is a very interesting and rapidly expanding literature that has
important antecedents (Wright, \citeyear{W21}) that were\linebreak[4]  influential in the
economics literature, and where Richardson and Sphitser have made major
contributions.
However, I saw the focus of my paper on an econometrics perspective on
instrumental variables, and there graphical models do not currently
play a major role.
It is an interesting question why economists have not felt that
graphical models have much to offer them.
\citet{Pea13} has also raised this question, and concludes somewhat
dismissively that: 
``economists
are still scared of graphs.'' He sees this as an ``educational
deficiency,'' and writes that
``This
educational impairment is the main factor that prevents economists from
appreciating
much of the recent progress in causal inference'' (Pearl, \citeyear{Pea13}, page 8).

My view on the lack of use in the econometrics literature on the
graphical models is more sanguine. I~see substantial evidence that as a
group economists are willing to adopt new methods from other
disciplines that are viewed as useful in practice. There are many
examples of this even within the area of causal inference. The rapid
adoption of the Rubin potential outcome approach starting in the early
1990s with Heckman (1990) and Manski (1990) is one, as is the by now
widespread use of matching and propensity score methods, and the
current boom in studies using methods associated with regression
discontinuity designs that were originally developed in the psychology
literature (see Cook, 2008, for a historical overview). In contrast,
the causal graphs have not caught on in economics. In my view a major
reason is that there have been few compelling applications of causal
graphs to social science questions where the causal-graph approach has
generated novel analyses or prevented researchers from making mistakes
that other frameworks might have encouraged them to make.
A second reason may be that some assumptions are not easy to
incorporate in the graphical approach. Monotonicity, which Swanson and
Hern\'an are particularly concerned with in their comments, and which
plays a key role in instrumental variables analyses, is difficult to
capture in a causal graph. See the discussion in Imbens and Rubin (\citeyear{ImbRub95}).

Let me flesh out the first part of this argument.
There are thousands of empirical studies in economics where researchers
use instrumental variables methods. Implicitly, they may have a causal
graph like Figure~1 in the main paper, or Figure 1(c) in the Shpitser comment,
in mind. Often there is considerable discussion in a particular
application whether the two key assumptions that there is no direct
effect of $Z_i$ on $Y_i^{\mathrm{obs}}$ (no arrow from $Z_i$ to $Y_i$,
and no confounding of the effect of $Z_i$ on $Y^{\mathrm{obs}}_i$ (no
unobserved common cause of $Z_i$ and $Y_i$) are plausible. In
observational studies in social science, both these assumptions tend to
be controversial. In this relatively simple setting, I do not see the
causal graphs as adding much to either the under\-standing of the
problem, or to the analyses.
Similarly, there are thousands of empirical studies in economics where
researchers use matching type methods based on the assumption of no
unmeasured confounders, and where implicitly they may have a causal
graph like Figure~1(b) in mind. Again, the assumptions underlying such
a graph are typically controversial and researchers often put in
substantial effort in arguing for the absence of unobserved
confounders. In this case, again I fail to see what using a
causal-graph approach would add in practice.
Now consider a more complicated setting such as the
``hypothetical longitudinal study represented by the causal graph
shown in Figure~2,'' in the comment by Shpitser, or Figure~1 in Pearl
(1995). Here, identification questions are substantially more complex,
and there is a strong case that the graph-based analyses have more to
contribute. However, I am concerned about the relevance of such
examples in social science settings. I~would like to see more
substantive, rather than hypothetical, applications where a graph such
as that in Figure~2 could be argued to capture the causal structure.
There are a large number of assumptions coded into such graphs, and
given the difficulty in practice to argue for the absences of one or
two arrows in instrumental-variables or no-unobserved-confounders
applications in social sciences, I worry that in practice it is
difficult to convince readers that such a causal graph fully captures
all important dependencies.
In other words, in social sciences applications a graph with many
excluded links may not be an attractive way of modeling dependence
structures. As Andrew Gelman writes on his blog in a discussion of
graphical models and potential outcomes, ``Nothing is zero, everything
matters to some extent'' (\cite{Gel}). Of course, instrumental
variables methods do also critically rely on the absence of particular
dependencies, but my point is that the larger graphical models such as
those in Figure~2 of the Shpitser comment or Figure~1 in Pearl (1995)
with many variables and many excluded links require researchers to
evaluate critically many more of those assumptions. The causal graph
methods appear to be more suited to answering the question whether
given a complex set of conditional independencies particular causal
effects are identified, whereas in my experience in many social science
applications researchers proceed by assessing a few conditional
independencies given which it is known particular effects are identified.

\section*{Swanson and Hern\'an:
``Think Globally, Act Globally: An
Epidemiologist'S Perspective on
Instrumental Variable Estimation''}

First of all, I want to commend Swanson and Hern\'an for their work on
improving the reporting the results of instrumental variables analyses
(Swanson and Hern\'an, \citeyear{S13}). Although many of their recommendations
such as the reporting of estimates of the proportion of compliers are
routinely followed in the economics literature (these estimates are
there often referred to as the first stage coefficients in the
two-stage-least-squares terminology), these practices had not made it
to the epidemiology literature, and their work will likely improve
practice there.
I am also glad to see that they do not attempt to defend the
homogeneity assumptions that would allow for point identification of
the ATE: it appears that there is growing consensus that such
assumptions are not realistic.
There are other areas where there is less agreement.
Swanson and Hern\'an take issue with the focus in the paper on the
local average treatment effect (LATE). Whereas Kitagawa felt LATEs were
``valuable pieces of information about causal effects'' (Kitagawa, page
359), Swanson and Hern\'an take the view that ``the LATE is not generally
relevant to epidemiological questions'' and propose to ``refocus on the
global ATE in the population of interest'' (Swanson and Hern\'an, page 371).

In my response to Swanson and Hern\'an, I want to make three points.
First, I want to correct the record concerning my position on
presenting estimates based on IV assumptions. Swanson and Hern\'an
summarize my position in terms of ``two options \ldots (1) present bounds
for the ATE, \ldots, or (2) present point estimates'' (pages 372--373) and then add
that ``of course \ldots  we can always do both.'' Swanson and Hern\'an
appear to have missed that presenting both
the bounds and the point estimate for the LATE (which is the same as
the point estimate for the ATE under homogeneity) was what I in fact
proposed (see also the comments by Kitagawa). One concern with the sole
focus on the ATE that Swanson and Hern\'an appear to favor, either
directly, or in combination with tighter bounds on outcomes, is that
one may discard relevant information.\vadjust{\goodbreak} Let me expand on comments in the
main paper in this regard. Consider the following two versions of an
artificial example with a dichotomous instrument, treatment and
outcome. Let $p_{zxy}$ be the population fraction of units with
$Z_i=z$, $X_i=x$, and $Y_i=y$, for $z,x,y\in\{0,1\}$. In the first
example, suppose $p^1_{000}=1/4$, $p^1_{001}=1/12$, $p^1_{010}=0$,
$p^1_{011}=0$, $p^1_{100}=1/24$, $p^1_{101}=7/24$, $p^1_{110}=7/24$,
$p^1_{111}=1/24$, and suppose these fractions are estimated precisely.
In this case, the fractions of compliers, nevertakers and alwaystakers
are 1$/$2, 1$/$2 and 0, the bounds on the ATE are $[-3/16,5/16]$, and the
point estimate of the LATE is $-$1$/$4. In the second example,
$p^2_{000}=1/6$, $p^2_{001}=1/6$, $p^2_{010}=0$, $p^2_{011}=0$,
$p^2_{100}=1/8$, $p^2_{101}=5/24$, $p^2_{110}=1/8$, $p^2_{111}=5/24$.
In this case,
the fractions of compliers, nevertakers and alwaystakers are again 1$/$2,
1$/$2 and 0, the bounds on the ATE are the same, $[-3/16,5/16]$, and the
point estimate of the LATE is now positive 1$/$4. Under the instrumental
variables assumptions, the bounds for the ATE are identical in the two
examples, but the LATEs are very different. In the first case, there is
evidence of a substantial negative effect for a subpopulation, whereas
in the second example one knows there is a subpopulation for which the
effect is substantial and positive. That would appear to potentially
lead to very different substantive conclusions. Simply reporting bounds
would miss these results.

In the second part of my response to Swanson and Hern\'an, I will
discuss more explicitly the concerns about external validity that are
implicit in the discussions of the relative merits of the overall
average effect (ATE) and the LATE.
Swanson and Hern\'an are interested in the ATE in the population of
interest, and then without explicitly saying so, assume that the study
population is representative for this population of interest.
Matters are rarely so clear cut in practice. The study sample need not
be a random sample from the population of interest because of
nonresponse, or the policy maker may be interested in the average
effect if the treatment were to be extended to a larger population at a
future date, or were to be offered on a voluntary basis to the general
population. What the population of future volunteers looks like may
well depend on the efficacy of the treatment according to the
statistical analysis. There are many examples where even in randomized
experiments the causal effects found for the study population did not
generalize to the population subsequently subject to the treatment.
Once one recognizes that even the study population may differ from the
population of interest much of the concern with the LATE that Swanson
and Hern\'an raise loses its force.
My position here is again essentially similar to the \citet{ShaCamCoo02} view on the primacy of internal validity over external validity.

In the third part of my response, I will make some comments on the
monotonicity assumption.
Swanson and Hern\'an present a generic example where the monotonicity
condition is likely to be violated, and argue that the instrument in
this example is one of the ``most commonly proposed instruments in
epidemiology.'' In fact, the example demonstrates how much there is to
be gained from a closer study of the earlier econometric literature, as it was discussed in the original paper on the LATE
(Example 2, page 472, Imbens  and Angrist, 1994); see also Section 5.3 in the current paper.
The generic example is as follows. The assignment of individuals to the
treatment is partly based on preferences of an administrator (physician
in the epidemiological version of the experiment). The assignment of
administrators to individuals is as good as random. Different
administrators may have different preferences on average, but it need
not be the case that the resulting instrument is monotone because the
set of individuals who would be assigned to the treatment by one
administrator need not be a proper subset of the set of individuals who
would be assigned to the treatment by a second administrator. That
setting also arises in applications of instrumental variables in legal
settings where the administrator may be a randomly assigned judge: see Aizer and Doyle  (2013) with an application in the criminal justice system, and
Dobbie and Song (2013) with an application to bankruptcy proceedings. It is important to distinguish such
settings from those where the instrument corresponds to an increase in
the incentive to participate, in which case the monotonicity assumption
is plausible. It is precisely by articulating explicitly these
assumptions and describing the role they play that we may be able to
avoid misleading decision-making efforts.


\renewcommand{\bibname}{Additional References}

\end{document}